\documentclass[11pt]{article}
\pdfoutput=1

\usepackage{amsmath,amssymb,hyperref}
\usepackage{fullpage}

\begin{document}

\title{A modified version of the Koide formula from flavor nonets\\
       in a scalar potential model and in a Yukawaon model}
\author{Zhengchen Liang\textsuperscript{a, b, *},
        Zheng Sun\textsuperscript{a, \dag}\\
        \textsuperscript{a}%
        \normalsize\textit{College of Physics, Sichuan University, Chengdu 610064, P. R. China}\\
        \textsuperscript{b}%
        \normalsize\textit{Department of Physics, Tsinghua University, Beijing 100084, P. R. China}\\
        \normalsize\textit{E-mail:}
        \textsuperscript{*}\texttt{lzc21@mails.tsinghua.edu.cn,}
        \textsuperscript{\dag}\texttt{sun\_ctp@scu.edu.cn}
       }
\date{}
\maketitle

\begin{abstract}
We present a modified version of the Koide formula from a scalar potential model or from a Yukawaon model, based on scalar fields set up in a nonet representation of the $\mathrm{SU}(3)$ flavor symmetry in the Standard Model.  The Koide's character, which involves the Standard Model fermion mass ratios, is derived from the vacuum expectation value of the nonet field in either model.  The scalar potential in the scalar potential model or the superpotential in the Yukawaon model is constructed with all terms invariant under symmetries.  The resulting Koide's character, which is modified by two effective parameters, can fit the experimental mass data of charged leptons, up quarks and down quarks.  It offers a natural interpretation of the Standard Model fermion mass spectrum.
\end{abstract}

\section{Introduction}

Patterns of the Standard Model (SM) fermion masses may suggest undiscovered new physics in the SM Yukawa sector.  Among various approaches to interpret the SM fermion mass spectrum, the Koide formula~\cite{Koide:1982si, Koide:1983a, Koide:1983b}
\begin{equation}
K = \frac{m_e + m_\mu + m_\tau}{\left ( \sqrt{m_e} + \sqrt{m_\mu} + \sqrt{m_\tau} \right )^2}
  = \frac{2}{3} \label{eq:1-01}
\end{equation}
exhibits a great consistency with the current experimental data~\cite{Patrignani:2016xqp, Tanabashi:2018oca, Zyla:2020zbs}.  From the data of charged lepton masses
\begin{align}
   m_e &= 0.5109989461 \pm 0.0000000031 \ \text{MeV} / c^2,\\
 m_\mu &= 105.6583745 \pm 0.0000024 \ \text{MeV} / c^2,\\
m_\tau &= 1776.86 \pm 0.12 \ \text{MeV} / c^2,
\end{align}
with their 1-$\sigma$ errors, the Koide's character $K$ is calculated to be
\begin{equation}
K = 0.6666605 \pm 0.0000068
  = \frac{2}{3} \times (0.999991 \pm 0.000010),
\end{equation}
which agrees with Eq.~\eqref{eq:1-01} within $10^{-5}$ precision and within the 1-$\sigma$ error.  In addition, when quantum electrodynamics (QED) radiative correction to charged lepton masses are taken into account, the value of $K$ deviates from $2 / 3$ by about $10^{- 3}$~\cite{Li:2006et, Xing:2006vk, Xing:2019vks}, which is $10^2$ times larger than the present experimental error.  To fix this deviation, the SM flavor symmetry is gauged, and the gauge interaction induces another radiative correction which may cancel the QED correction~\cite{Sumino:2008hu, Sumino:2008hy, Sumino:2009bt, Koide:2012kn, Koide:2014doa, Koide:2016bte, Koide:2016qeq}.  So the Koide formula Eq.~\eqref{eq:1-01} holds for both the pole masses and the running masses of charged leptons.  The Koide formula can be geometrically interpreted as the angle $\theta = \pi / 4$ between the two vectors $\left ( \sqrt{m_e}, \sqrt{m_\mu}, \sqrt{m_\tau} \right )$ and $(1, 1, 1)$ in three dimensions~\cite{Foot:1994yn}.  Empirical extensions of the Koide formula to masses of quarks and neutrinos are also conjectured in literature~\cite{Esposito:1995bw, Li:2005rp, Rivero:2005vj, Gerard:2005ad, Guo:2007rn, Rodejohann:2011jj, Kartavtsev:2011jt, Cao:2012un, Zenczykowski:2012fg, Zenczykowski:2013bb, Gao:2015xnv, Huang:2016ocs, Huang:2016sht}.
  
Early proposals to explain the physical origin of the Koide formula include radiative mass generation models, Froggatt-Nielsen type models~\cite{Froggatt:1978nt} and seesaw-type models, with assumptions of discrete flavor symmetries, democratic mixing or tribimaximal mixing~\cite{Koide:1989ds, Koide:1989jq, Koide:1992vs, Koide:1993da, Koide:1994wu, Koide:1995ie, Koide:1995xk, Koide:1999mx, Koide:2000zi, Koide:2005nv, Koide:2005za, Koide:2005ep, Mohapatra:2006gs, Koide:2006dn, Ma:2006ht, Koide:2006vs, Koide:2007kw, Koide:2007sr, Koide:2007eu, Koide:2010vu}.  A more recent development is known as the Yukawaon model~\cite{Haba:2008wr, Koide:2008eq, Koide:2008ey, Koide:2008tr, Koide:2010zz}, in which the SM Yukawa coupling constant for charged leptons is induced by the vacuum expectation value (VEV) of a 9-component Hermitian matrix-valued scalar field $Y^i_j$, with flavor indices $i, j = 1, 2, 3$.  The nonet field $Y$ beyond SM, named the Yukawaon, can be set up in a $\mathbf{3} \otimes \mathbf{3}^* = \mathbf{8} \oplus \mathbf{1}$ representation of the $\mathrm{SU(3)}$ flavor symmetry,%
\footnote{The earliest usage of the term ``Yukawaon'' appears in~\cite{Koide:2008qm} referring to a sextet field in $\mathbf{3} \odot \mathbf{3} = \mathbf{5} \oplus \mathbf{1}$ of the $\mathrm{SO(3)}$ flavor symmetry, and in~\cite{Koide:2008tr} referring to a nonet field in $\mathbf{3} \otimes \mathbf{3}^* = \mathbf{8} \oplus \mathbf{1}$ of the $\mathrm{SU(3)}$ flavor symmetry.  For simplicity, here we only discuss nonet Yukawaons.}
which makes charged leptons anomaly free in the $\mathrm{SU(3)}$~\cite{Koide:2013eca}.  Yukawaon models for up quarks and down quarks can also be build in a similar way by introducing another two Yukawaons.  The SM Yukawa coupling terms are then replaced by dimension-five effective operators involving these Yukawaons.  We introduce three sectors of Yukawaons $Y^{(a)}$ with sector indices $a = e, u, d$, and arrange
\begin{equation}
\mathcal{L}^{(5)}_{\text{Yukawaon}} = - \frac{y_0}{\Lambda}
                                        \left ( \bar{l}_{L i} Y^{(e) i}_j H e^j_R
                                                + \bar{q}_{L i} Y^{(u) i}_j \tilde{H} u^j_R
                                                + \bar{q}_{L i} Y^{(d) i}_j H d^j_R
                                        \right )
                                      + \text{c.c.}, \label{eq:1-02}
\end{equation}
for the left-handed leptons $l^i_L = (\nu^i_L, e^i_L)^{\mathrm{T}}$, the right-handed charged leptons $e^i_R$, the left-handed quarks $q^i_L = (u^i_L, d^i_L)^{\mathrm{T}}$, the right-handed up quarks $u^i_R$, the right-handed down quarks $d^i_R$, the Higgs field $H = (H^+, H^0)^{\mathrm{T}}$ and its charge conjugation $\tilde{H} = \epsilon H^*$, with a dimensionless coefficient $y_0$ and an energy scale $\Lambda \gg m_W$.  The $\mathrm{SU}(3)$ flavor symmetries can be extended to $\mathrm{U}(3)$'s which include $\mathrm{U}(1)$'s corresponding to the conserved lepton and baryon numbers, and Eq.~\eqref{eq:1-02} is also invariant under the $\mathrm{U}(3)$'s by setting all $\mathrm{U}(1)$ charges of Yukawaons to $0$.  Below the scale $\Lambda$, the Yukawaons acquire non-zero VEV's in a quadratic form
\begin{equation}
\left \langle Y^{(a)} \right \rangle
\propto
\left \langle \Phi^{(a)} \right \rangle \left \langle \Phi^{(a)} \right \rangle,
\end{equation}
where the new $\mathrm{SU}(3)$ flavor nonet fields $\Phi^{(a)}$ are named the ur-Yukawaons.  Such VEV relations can be obtained from the F-flatness condition with a superpotential, e.g.~\cite{Koide:2008tr},
\begin{align}
W_{\text{Yukawaon}} &= W_0 + W^{(\Phi)}(\Phi^{(a)}, \phi_i), \label{eq:1-03}\\
                W_0 &= W^{(e)}_0 + W^{(u)}_0 + W^{(d)}_0,\\
          W^{(a)}_0 &= \lambda^{(a)}_A \operatorname{Tr} \left [ \Phi^{(a)} \Phi^{(a)} A^{(a)} \right ]
                       + \mu^{(a)}_A \operatorname{Tr} \left [ Y^{(a)} A^{(a)} \right ],
\end{align}
where another set of nonet fields $A^{(a)}$ are introduced.  The $W^{(\Phi)}$ part of Eq.~\eqref{eq:1-03} may contain more chiral superfields $\phi_i$ to help the stabilization of $\Phi$'s at their VEV's.  At an SUSY vacuum satisfying the F-flatness condition, we have
\begin{equation}
\left \langle Y^{(a)} \right \rangle = - \frac{\lambda^{(a)}_A}{\mu^{(a)}_A}
                                         \left \langle \Phi^{(a)} \right \rangle
                                         \left \langle \Phi^{(a)} \right \rangle, \quad
\left \langle A^{(a)} \right \rangle = 0, \quad
\left \langle \partial_{\Phi^{(a)}} W^{(\Phi)} \right \rangle = 0, \quad
\left \langle \partial_{\phi_i} W^{(\Phi)} \right \rangle = 0. \label{eq:1-04}
\end{equation}
Non-zero VEV's of $\Phi$'s and $Y$'s spontaneously break the $\mathrm{SU}(3)$ flavor symmetry in each sector, and Eq.~\eqref{eq:1-02} becomes the SM Yukawa coupling terms with the effective Yukawa coupling coefficients
\begin{equation}
y^{(a) i}_j = \frac{y_0}{\Lambda} \left \langle Y^{(a) i}_j \right \rangle
            = - \frac{y_0 \lambda^{(a)}_A}{\Lambda \mu^{(a)}_A}
                \left \langle \Phi^{(a) i}_k \right \rangle
                \left \langle \Phi^{(a) k}_j \right \rangle.
\end{equation}
Then the VEV of the Higgs doublet $\langle H \rangle = (0, v / \sqrt{2})^{\mathrm{T}}$ gives the the SM fermion mass matrices 
\begin{equation}
M^{(a) i}_j = \frac{v}{\sqrt{2}} y^{(a) i}_j
            = - \frac{y_0 \lambda^{(a)}_A v}{\sqrt{2} \Lambda \mu^{(a)}_A}
                \left \langle \Phi^{(a) i}_k \right \rangle
                \left \langle \Phi^{(a) k}_j \right \rangle. \label{eq:1-05}
\end{equation}
For charged leptons, it is known from experiments with high precision that the mass matrix is diagonal, i.e.\ $M^{(e)} = \operatorname{diag} (m_e, m_\mu, m_\tau)$.  The above effective mechanism gives the Koide's character
\begin{equation}
K = \frac{m_e + m_\mu + m_\tau}{\left ( \sqrt{m_e} + \sqrt{m_\mu} + \sqrt{m_\tau} \right )^2}
  = \frac{[\Phi \Phi]}{[\Phi]^2}. \label{eq:1-06}
\end{equation}
For convenience, we use the notation $[X] \equiv \operatorname{Tr}[X]$ where $X$ can be any monomial of nonet fields, and write $\Phi$ for $\langle \Phi^{(e)} \rangle = \left \langle \Phi \right \rangle$ as long as it does not cause any ambiguity.  For up quarks and down quarks, their masses are given by the singular values of the mass matrices.  Although the quark mass matrices are not diagonal, the experimental data of quark masses and the Cabibbo-Kobayashi-Maskawa (CKM) matrix can always be fitted with the Yukawa coupling matrices being Hermitian.  So the Yukawaons and ur-Yukawaons in up quark and down quark sectors can also be assumed to be Hermitian matrix-valued scalar fields, which can be diagonalized by unitary similarity transformations.  We still have the similar mass formulas expressed by the VEV's of $\Phi$'s:
\begin{align}
  K_{\text{up quarks}} &= \frac{m_u + m_c + m_t}{\left ( \sqrt{m_u} + \sqrt{m_c} + \sqrt{m_t} \right )^2}
                        = \frac{\left [ \Phi^{(u)} \Phi^{(u)} \right]}{\left [ \Phi^{(u)} \right ]^2}, \label{eq:1-07}\\
K_{\text{down quarks}} &= \frac{m_d + m_s + m_b}{\left ( \sqrt{m_d} + \sqrt{m_s} + \sqrt{m_b} \right )^2}
                        = \frac{\left [ \Phi^{(d)} \Phi^{(d)} \right]}{\left [ \Phi^{(d)} \right ]^2}. \label{eq:1-08}
\end{align}
With a properly arranged superpotential, the VEV's of $\Phi$'s give the values of $K$'s fitting the experimental data.  Following this routine, different models have been build to reproduce the Koide formula Eq.~\eqref{eq:1-01}, its analogues for quark masses and neutrino masses, and the fermion mixing matrices~\cite{Koide:2008qm, Koide:2008sj, Koide:2008he, Koide:2008kw, Koide:2009zz, Koide:2010np, Koide:2010hp, Nishiura:2010rt, Koide:2012zz, Koide:2011wj, Koide:2012fw, Koide:2012ji, Koide:2013ie, Koide:2014nxa, Koide:2014oxa, Koide:2015ura, Koide:2015hya, Koide:2015ype, Koide:2017lan, Koide:2018fsj}.

The usage of flavor nonets to explain the Koide formula can also be traced back to some early works on seesaw-type models~\cite{Koide:1989jq, Koide:1992vs, Koide:1994wu, Koide:1995ie, Koide:1995xk, Koide:2007sr, Koide:2010vu}, in which a flavor nonet scalar $\Phi^{(e)} = \Phi$, a flavor singlet scalar $S^{(e)} = S$, and new heavy fermions $L^{(l)}_L = L_L = (N_L, F_L)^{\mathrm{T}}$ and $F^{(e)}_R = F_R$ are introduced for the charged lepton sector.  Lepton masses are generated from the VEV's of $\Phi$, $S$ and the SM Higgs field $H$.  From the dimension-five effective operators
\begin{equation}
\mathcal{L}^{(5)}_{\text{seesaw}} = - \frac{y_0}{\Lambda}
                                      \left ( \bar{l}_{L i} \Phi^i_j H F^j_R
                                              + \bar{L}_{L i} \Phi^i_j H e^j_R
                                              + \bar{L}_{L i} S H F^i_R
                                      \right )
                                    + \text{c.c.}, \label{eq:1-09}
\end{equation}
we obtain the see-saw type mass terms
\begin{equation}
\begin{split}
\mathcal{L}_{\text{seesaw}} &= - \bar{e}_L m_L F_R
                               - \bar{F}_L m_R e_R
                               - \bar{F}_L M_F F_R
                               + \text{c.c.}\\
                            &= - \begin{pmatrix}
                                 \bar{e}_{L i} & \bar{F}_{L j}
                                 \end{pmatrix}
                                 \begin{pmatrix}
                                 0         & m^i_{L l}\\
                                 m^j_{R k} & M_F \delta^j_l
                                 \end{pmatrix}
                                 \begin{pmatrix}
                                 e^k_R\\
                                 F^l_R
                                 \end{pmatrix}
                                + \text{c.c.},
\end{split} \label{eq:1-10}
\end{equation}
with effective mass parameters
\begin{equation}
m^i_{L j} = m^i_{R j}
          = \frac{y_0 v}{\sqrt{2} \Lambda} \Phi^i_j, \quad
M_F = \frac{y_0 v}{\sqrt{2} \Lambda} S.
\end{equation}
Mass eigenvalues are found by a singular value decomposition (SVD) of the mass matrix in Eq.~\eqref{eq:1-10}.  Assuming that the VEV of $S$ is much larger than the VEV of any component of $\Phi$, or $M_F \gg \lVert m_L \rVert = \lVert m_R \rVert$, the singular values are approximately identical to the magnitudes of the eigenvalues of the mass matrix.  The heavy mass eigenstates are almost aligned to $F_L$ and $F_R$, with their mass $M \approx M_F$.  The light mass eigenstates, which correspond to the SM charged leptons, have the seesaw mass matrix
\begin{equation}
M^{(e) i}_j \approx m^i_{L k} M_F^{-1} m^k_{R j}
            = \frac{y_0 v}{\sqrt{2} \Lambda S} \Phi^i_k \Phi^k_j.
\end{equation}
Similarly to  Eq.~\eqref{eq:1-05} in the Yukawaon model, here $M^{(e)}$ is also quadratic in $\Phi$.  Thus the same expression of the Koide's character Eq.~\eqref{eq:1-06} is obtained.  Expressions Eq.~\eqref{eq:1-07} for up quarks and Eq.~\eqref{eq:1-08} for down quarks can also be derived in a similar seesaw-type setup.  With a properly arranged scalar potential $V(\Phi)$, the VEV of $\Phi$ gives the value of $K$ fitting the experimental data.  Such a model is also referred to as the scalar potential model, with emphasis on the scalar potential $V(\Phi)$ instead of the seesaw mechanism which leads to the mass matrix quadratic in $\Phi$.

However, in previous versions of scalar potential models or Yukawaon models, the scalar potential $V$ or the superpotential $W$ is incomplete, i.e., it does not include all possible terms respecting $\mathrm{SU}(3)$.  The missing $\mathrm{SU}(3)$-invariant terms in $V$ or $W$ must be unnaturally fine-tuned to zero.  A recent approach~\cite{Koide:2017lrf} considers all $\mathrm{SU}(3)$-invariant terms in the scalar potential.  Besides the original Koide formula, a second formula on charged lepton masses is proposed.  But some vital mistakes in the derivation actually invalidate the result, although the second formula can be successfully derived in other Yukawaon models with a suitable superpotential $W$~\cite{Koide:2009hn, Koide:2018gdm}.  As we are going to show, the complete scalar potential instead leads to a modified version of the Koide formula.

In this work, we take the scalar potential $V$ to include all $\mathrm{SU}(3)$-invariant terms up to quartic.  A $\mathbb{Z}_2$ symmetry is imposed to eliminate linear and cubic terms, and a Higgs-like quadratic term is assumed to generate the non-zero VEV of $\Phi$.  The equations for a stationary point lead to a modified version of the Koide formula.  The modified formula is then reproduced in a Yukawaon model.  The nonet $\Phi$ is promoted to a nonet chiral superfield, and two additional chiral superfields are introduced.  The superpotential $W$ includes all $\mathrm{SU}(3)$-invariant terms up to cubic.  An R-symmetry is imposed to further restrict the form of $W$, and a small R-symmetry breaking term is introduced to generate the non-zero VEV of $\Phi$.  The F-flatness condition for a SUSY vacuum leads to the same VEV relations and the same modified formula as the result from the scalar potential model.  In both models, the Koide's character $K$ is modified by two effective parameters.  The modified range of $K$ covers all possible values of $K$ for charged leptons, up quarks and down quarks.  It offers a natural interpretation of SM fermion mass spectrum.

The rest part of this paper is arranged as following.  In Section \ref{sec:2}, we derive the modified version of the Koide formula from a scalar potential with all $\mathrm{SU}(3)$-invariant terms.  In Section \ref{sec:3}, from a superpotential constructed with the $\mathrm{SU}(3)$ flavor symmetry and an R-symmetry, we derived the VEV relations and the modified formula in the Yukawaon model, which is in agreement with the result from the scalar potential model.  In Section \ref{sec:4}, we make some concluding remarks on the possible implications from the modified formula.

\section{The modified formula from a scalar potential model} \label{sec:2}

Following the previous approach~\cite{Koide:2017lrf}, we consider the scalar potential $V$ with all $\mathrm{SU}(3)$-invariant terms of the nonet scalar field $\Phi$ presented as the octet $\Phi_8$ and the singlet $[\Phi]$.  A renormalizable $V$ contains terms only up to quartic.  Besides the $\mathrm{SU}(3)$ flavor symmetry, we impose a $\mathbb{Z}_2$ symmetry under which both $\Phi_8$ and $[\Phi]$ are odd.  Thus only quadratic and quartic terms are allowed in $V$.  For quartic terms, we have
\begin{equation}
V_1 = a_0 [\Phi_8 \Phi_8 \Phi_8 \Phi_8]
      + a_{02} [\Phi_8 \Phi_8] [\Phi_8 \Phi_8]
      + a_1 [\Phi_8 \Phi_8 \Phi_8] [\Phi]
      + a_2 [\Phi_8 \Phi_8] [\Phi]^2
      + a_4 [\Phi]^4.
\end{equation}
Note that the dimension-five operators in Eq.~\eqref{eq:1-09} are also $\mathbb{Z}_2$-invariant by letting the Higgs field $H$ and the singlet $S$ to be odd and all other SM fields to be neutral.  For quadratic terms, we assume that they combine to a Higgs-like negative mass-square term to generate the non-zero VEV of $\Phi$:
\begin{equation}
V_2 = - \mu^2 [\Phi \Phi]
    = - \mu^2 [\Phi_8 \Phi_8]
      - \frac{1}{3} \mu^2 [\Phi]^2.
\end{equation}
Both $V_1$ and $V_2$ combine into the full scalar potential
\begin{equation}
V = V_1 + V_2.
\end{equation}
The octet and the singlet appear in the Clebsch-Gorden series of $\mathrm{SU}(3)$ representations $\mathbf{3} \otimes \mathbf{3}^* = \mathbf{8} \oplus \mathbf{1}$.  They combine to the nonet scalar field
\begin{equation}
\Phi = \Phi_8 + \frac{1}{3} [\Phi] \mathbf{I}_{3 \times 3}
     = \Phi_8^a t^a + [\Phi] t^0, \label{eq:2-01}
\end{equation}
where we denote $t^0 = \frac{1}{3} \mathbf{I}_{3 \times 3}$.  The matrices $t^a$'s, where $a = 1, \dotsc , 8$, are eight generators of the Lie algebra $\mathfrak{su}(3)$.  Altogether $t^0$ and $t^a$'s give nine generators of $\mathfrak{u}(3) \cong \mathfrak{su}(3) \times \mathfrak{u}(1)$, and the nine real scalar fields $\{ \Phi_8^a, [\Phi] \}$ give the Hermitian matrix-valued field $\Phi \in \mathfrak{u}(3)$.  Replacing $\Phi_8$ with $\Phi$ and $[\Phi]$, the following identities can be derived:
\begin{align}
           {} [\Phi_8 \Phi_8] &= [\Phi \Phi]
                                 - \frac{1}{3} [\Phi]^2,\\
       [\Phi_8 \Phi_8 \Phi_8] &= [\Phi \Phi \Phi]
                                 - [\Phi \Phi] [\Phi]
                                 + \frac{2}{9} [\Phi]^4,\\
[\Phi_8 \Phi_8 \Phi_8 \Phi_8] &= [\Phi \Phi \Phi \Phi]
                                 - \frac{4}{3} [\Phi \Phi \Phi] [\Phi]
                                 + \frac{2}{3} [\Phi \Phi] [\Phi]^2
                                 - \frac{1}{9} [\Phi]^4.
\end{align}
The scalar potential $V$ is then recast in terms of $[\Phi]$, $[\Phi \Phi]$, $[\Phi \Phi \Phi]$ and $[\Phi \Phi \Phi \Phi]$:
\begin{equation}
\begin{split}
V = \mbox{} & - \mu^2 [\Phi \Phi]
              + a_0 [\Phi \Phi \Phi \Phi]
              + a_{02} [\Phi \Phi] [\Phi \Phi]
              + \left ( - \frac{4}{3} a_0
                        + a_1
                \right ) [\Phi \Phi \Phi] [\Phi]\\
            & + \left ( \frac{2}{3} a_0
                        - \frac{2}{3} a_{02}
                        - a_1
                        + a_2
                \right ) [\Phi \Phi] [\Phi]^2
              + \left ( - \frac{1}{9} a_0
                        + \frac{1}{9} a_{02}
                        + \frac{2}{9} a_1
                        - \dfrac{1}{3} a_2
                        + a_4
                \right ) [\Phi]^4.
\end{split}
\end{equation}
With proper choices of parameters $a_0$, $a_{02}$, $a_1$, $a_2$, $a_4$ and $\mu$, we expect that non-zero VEV's of $\Phi_8^a$ and $[\Phi]$ can be obtained to give the SM fermion masses through Eq.~\eqref{eq:1-05}.

The vacuum is a stationary point of $V$, i.e., the VEV's of the first derivatives of $V$ with respect to $\Phi_8^a$ and $[\Phi]$ must vanish:
\begin{equation}
\partial_{\Phi_8^a} V = \partial_{[\Phi]} V = 0. \label{eq:2-02}
\end{equation}
Since $V$ is a polynomial function of $\Phi_8^a$ and $[\Phi]$, it is also a holomorphic function if $\Phi_8^a$ and $[\Phi]$ are viewed as complex variables.  Eq.~\eqref{eq:2-02} is then equivalent to the same set of equations complexified with its solutions restricted to real numbers.  Known that the complexification of $\mathfrak{u}(3)$ is isomorphic to $\mathfrak{gl}(3, \mathbb{C})$ viewed as a complex Lie algebra, i.e., $\mathfrak{u}(3)_\mathbb{C} \cong \mathfrak{gl}(3, \mathbb{C})$, the linear map from nine complex fields $\Phi_8^a$ and $[\Phi]$ to the complex matrix $\Phi \in \mathfrak{gl}(3, \mathbb{C}) = \mathbb{C}^{3 \times 3}$ is bijective.  So $V$ can also be interpreted as a function of the nine independent complex matrix components of $\Phi$, and Eq.~\eqref{eq:2-02} is equivalent to
\begin{equation}
\partial_\Phi V = \begin{pmatrix}
                  \partial_{\Phi^1_1} V & \partial_{\Phi^1_2} V & \partial_{\Phi^1_3} V\\
                  \partial_{\Phi^2_1} V & \partial_{\Phi^2_2} V & \partial_{\Phi^2_3} V\\
                  \partial_{\Phi^3_1} V & \partial_{\Phi^3_2} V & \partial_{\Phi^3_3} V\\
                  \end{pmatrix}
                = 0.
\end{equation}

Applying the identities for matrix derivatives
\begin{equation}
\partial_\Phi [\Phi^n] = n \Phi^{n-1}, \quad
\partial_\Phi [\Phi] = \mathbf{I}_{3 \times 3},
\end{equation}
we have
\begin{equation}
\begin{split}
0 = \partial_\Phi V
  = \mbox{} & - 2 \mu^2 \Phi
              + 4 a_0 \Phi \Phi \Phi
              + 4 a_{02} [\Phi \Phi] \Phi
              + \left ( - \frac{4}{3} a_0
                        + a_1
                \right )
                (3 [\Phi] \Phi \Phi
                 + [\Phi \Phi \Phi] \mathbf{I}_{3 \times 3}
                )\\
            & + 2 \left ( \frac{2}{3} a_0
                          - \frac{2}{3} a_{02}
                          - a_1
                          + a_2
                  \right )
                  \left ( [\Phi]^2 \Phi
                          + [\Phi \Phi] [\Phi] \mathbf{I}_{3 \times 3}
                  \right )\\
            & + 4 \left ( - \frac{1}{9} a_0
                          + \frac{1}{9} a_{02}
                          + \frac{2}{9} a_1
                          - \dfrac{1}{3} a_2
                          + a_4
                  \right ) [\Phi]^3 \mathbf{I}_{3 \times 3}.
\end{split} \label{eq:2-03}
\end{equation}
Notice that for any $3 \times 3$ complex matrix $\Phi \in \mathbb{C}^{3 \times 3}$, the following identities can be verified by straightforward calculation:
\begin{align}
\Phi \Phi \Phi   &= [\Phi] \Phi \Phi
                    + \frac{1}{2} \left ( [\Phi \Phi]
                                          - [\Phi]^2
                                  \right ) \Phi
                    + \det(\Phi) \mathbf{I}_{3 \times 3},\\
[\Phi \Phi \Phi] &= \frac{3}{2} [\Phi \Phi] [\Phi]
                    - \frac{1}{2} [\Phi]^3
                    + 3 \det(\Phi).
\end{align}
Using these identities, the vacuum equation Eq.~\eqref{eq:2-03} can be rearranged into
\begin{equation}
\begin{split}
0 = \partial_\Phi V
  = \mbox{} & 3 a_1 [\Phi] \Phi \Phi
              + 2 \left ( - \mu_1^2
                          + (a_0 + 2 a_{02}) [\Phi \Phi]
                          + \left ( - \frac{1}{3} a_0
                                    - \frac{2}{3} a_{02}
                                    - a_1
                                    + a_2
                            \right ) [\Phi]^2
                  \right ) \Phi\\
            & + 2 \left ( \left ( - \frac{1}{3} a_0
                                  - \frac{2}{3} a_{02}
                                  - \frac{1}{4} a_1
                                  + a_2
                          \right ) [\Phi \Phi] [\Phi] \right.\\
            & \hphantom{+ 2 \left ( \vphantom{\frac{1}{3}} \right.}
                  \left. \mbox{}
                          + \left ( \frac{1}{9} a_0
                                    + \frac{2}{9} a_{02}
                                    + \frac{7}{36} a_1
                                    - \frac{2}{3} a_2
                                    + 2 a_4
                            \right ) [\Phi]^3
                          + \frac{3}{2} a_1 \det(\Phi)
                  \right ) \mathbf{I}_{3 \times 3}.
\end{split} \label{eq:2-04}
\end{equation}

The VEV's of $\Phi \Phi$, $\Phi$ and $\mathbf{I}_{3 \times 3}$ are linearly independent for a non-zero VEV of $\Phi$ without fine-tuning.  So their corresponding coefficients in Eq.~\eqref{eq:2-04} must vanish:
\begin{align}
0 &=  3 a_1 [\Phi], \label{eq:2-05}\\
0 &= - \mu^2
     + (a_0 + 2 a_{02}) [\Phi \Phi]
     + \left ( - \frac{1}{3} a_0
               - \frac{2}{3} a_{02}
               - a_1
               + a_2
       \right ) [\Phi]^2, \label{eq:2-06}\\
\begin{split}
0 &= \left ( - \frac{1}{3} a_0
             - \frac{2}{3} a_{02}
             - \frac{1}{4} a_1
             + a_2
     \right ) [\Phi \Phi] [\Phi]\\
  & \hphantom{= \mbox{}}
     + \left ( \frac{1}{9} a_0
               + \frac{2}{9} a_{02}
               + \frac{7}{36} a_1
               - \frac{2}{3} a_2
               + 2 a_4
       \right ) [\Phi]^3
     + \frac{3}{2} a_1 \det(\Phi). \label{eq:2-07}
\end{split}
\end{align}
Assuming $[\Phi]$ gets a non-zero VEV, Eq.~\eqref{eq:2-05} gives
\begin{equation}
a_1 = 0,
\end{equation}
which implies that the term $a_1 [\Phi_8 \Phi_8 \Phi_8] [\Phi]$ must vanish in the scalar potential $V$ in order to get a non-zero VEV of $\Phi$.  Then Eq.~\eqref{eq:2-07} becomes
\begin{equation}
0 = \left ( - \frac{1}{3} a_0
            - \frac{2}{3} a_{02}
            + a_2
    \right ) [\Phi \Phi]
    + \left ( \frac{1}{9} a_0
              + \frac{2}{9} a_{02}
              - \frac{2}{3} a_2
              + 2 a_4
      \right ) [\Phi]^2,
\end{equation}
which gives a modified version of the Koide formula:
\begin{equation}
K = \frac{[\Phi \Phi]}{[\Phi]^2}
  = \frac{2}{3} \times \frac{(a_0 + 2 a_{02}) / 6 - a_2 + 3 a_4}{(a_0 + 2 a_{02}) / 3 - a_2}
  = \frac{2}{3} \times \left ( 1 - \frac{a_0 + 2 a_{02} - 18 a_4}{2(a_0 + 2 a_{02} - 3 a_2)} \right ). \label{eq:2-08}
\end{equation}
Note that the term involving $\det(\Phi)$ disappears as $a_1$ becomes zero, thus the derivation of the second formula in~\cite{Koide:2017lrf} is invalid.  Finally Eq.~\eqref{eq:2-06} gives
\begin{equation}
\mu^2 = (a_0 + 2 a_{02}) [\Phi \Phi]
        - \left ( \frac{1}{3} a_0
                 + \frac{2}{3} a_{02}
                 - a_2
          \right ) [\Phi]^2
      = \left ( (a_0 + 2 a_{02})
                \left( K - \frac{1}{3} \right)
                + a_2
        \right ) [\Phi]^2. \label{eq:2-09}
\end{equation}
The coefficient of $[\Phi]^2$ is zero only if parameters are fine-tuned to satisfy $a_2^2 = 2 a_4 (a_0 + 2 a_{02})$.  Without such fine-tuning, a non-zero VEV of $\Phi$ requires a non-zero $\mu$.

The modified formula Eq.~\eqref{eq:2-08} contains four free parameters $a_0$, $a_{02}$, $a_2$ and $a_4$.  Among them, $a_0$ and $a_{02}$ appears in the fixed combination $a_0 + 2 a_{02}$.  In addition, $K$ only depends on ratios of the parameters.  Hence there are only two effective parameters in the modified formula.  Although the parameter $a_1$ is allowed by symmetries, it has to be tuned to zero in order to generate a non-zero VEV of $[\Phi]$.  We may further explore other models which lead to the modified formula with less tuning.  Such a model is constructed in the following SUSY scenario.

\section{The modified formula from a Yukawaon model} \label{sec:3}

Enlightened by previous derivation of the Koide formula from Yukawaon models, such as the work of~\cite{Koide:2008tr}, we consider a superpotential $W(z_i)$ of chiral superfields $\{ z_i \}$ in the Wess-Zumino model.  In addition to the $\mathrm{SU}(3)$ flavor symmetry, we impose a $\mathrm{U}(1)$ R-symmetry under which $W$ has R-charge $2$.  We introduce two $\mathrm{SU}(3)$ singlet chiral superfields $\phi'_1$ and $\phi'_2$, both with R-charge $1$.  The nonet scalar field $\Phi$ in Eq.~\eqref{eq:2-01} is complexified to $\Phi \in \mathfrak{u}(3)_\mathbb{C} \cong \mathfrak{gl}(3, \mathbb{C}) = \mathbb{C}^{3 \times 3}$, and then promoted to a nonet chiral superfield with R-charge $1/2$.  The superpotential
\begin{equation}
W_1 = \frac{1}{2}
      \begin{pmatrix}
      \phi'_1 & \phi'_2
      \end{pmatrix}
      \begin{pmatrix}
      \mu'_{11} & \mu'_{12}\\
      \mu'_{12} & \mu'_{22}
      \end{pmatrix}
      \begin{pmatrix}
      \phi'_1\\
      \phi'_2
      \end{pmatrix}
      + \begin{pmatrix}
        \phi'_1 & \phi'_2
        \end{pmatrix}
        \begin{pmatrix}
        b'_{11} & b'_{12}\\
        b'_{21} & b'_{22}
        \end{pmatrix}
        \begin{pmatrix}
        [\Phi_8 \Phi_8]\\
        [\Phi]^2
        \end{pmatrix}
\end{equation}
includes all renormalizable terms respecting the $\mathrm{SU}(3)$ flavor symmetry and the R-symmetry.  In addition, we introduce
\begin{equation}
W_2 = \mu_0 [\Phi \Phi]
    = \mu_0 [\Phi_8 \Phi_8]
      + \frac{1}{3} \mu_0 [\Phi]^2,
\end{equation}
which slightly breaks the R-symmetry with a small $\mu_0$.  This small R-symmetry breaking may be resolved by promoting $\mu_0$ to be an R-charge $1/2$ field, which gets a VEV from an extra sector beyond the scope of our current discussion.  Both $W_1$ and $W_2$ combine into the superpotential
\begin{equation}
W(\Phi, \phi'_1, \phi'_2) = W_1 + W_2,
\end{equation}
which contributes to the full superpotential as the $W^{(\Phi)}$ part of Eq.~\eqref{eq:1-03}.

With a redefinition of $\{ \phi'_1, \phi'_2 \}$, it is possible to make the quadratic term in $W_1$ off-diagonal.  The coefficient matrix
\begin{equation}
M = \begin{pmatrix}
    \mu'_{11} & \mu'_{12} \\
    \mu'_{12} & \mu'_{22}
    \end{pmatrix}
\end{equation}
is a $2 \times 2$ complex symmetric matrix.  It can be first diagonalized by a Autonne-Takagi factorization~\cite{Autonne:1915a, Takagi:1924a}
\begin{equation}
U^\text{T} M U = D
               = \begin{pmatrix}
                 \mu_1 & 0 \\
                 0     & \mu_2
                 \end{pmatrix}
\end{equation}
with a unitary matrix $U$.  One can then define a transformation matrix
\begin{equation}
P = U D^{- \frac{1}{2}} O C, \quad
\text{where} \
O \in \mathrm{O}(2), \
C = \sqrt{\frac{\mu_3}{2}}
    \begin{pmatrix}
    1 & 1 \\
    i & - i
    \end{pmatrix},
\end{equation}
so that
\begin{equation}
P^\text{T} M P = \begin{pmatrix}
                 0     & \mu_3 \\
                 \mu_3 & 0
                 \end{pmatrix}.
\end{equation}
Note that $P$ is not unique because $O$ can be an arbitrary orthogonal matrix.  A simple choice is
\begin{equation}
\begin{gathered}
P = \begin{pmatrix}
    \mu_\pm     & - \mu_{22}\\
    - \mu_{11}  & \mu_\pm
    \end{pmatrix}
    \begin{pmatrix}
    \frac{\mu_3}{2 \Delta} & 0\\
    0                      & \frac{1}{\mu_\pm}
    \end{pmatrix}
  = \begin{pmatrix}
    \frac{\mu_3 \mu_\pm}{2 \Delta}     & \frac{- \mu_{22}}{\mu_\pm}\\
    - \frac{\mu_3 \mu_{11}}{2 \Delta} & 1
    \end{pmatrix},\\
\text{where} \
\Delta = \mu_{12}^2 - \mu_{11} \mu_{22}, \
\mu_\pm = \mu_{12} \pm \sqrt{\Delta}.
\end{gathered}
\end{equation}
With the redefinition
\begin{equation}
\begin{pmatrix}
\phi'_1\\
\phi'_2
\end{pmatrix}
= P 
  \begin{pmatrix}
  \phi_1\\
  \phi_2
  \end{pmatrix}, \quad
\begin{pmatrix}
b'_{11} & b'_{12}\\
b'_{21} & b'_{22}
\end{pmatrix}
= \left ( P^\text{T} \right )^{- 1}
  \begin{pmatrix}
  b_{11} & b_{12}\\
  b_{21} & b_{22}
  \end{pmatrix},
\end{equation}
and replacing $\Phi_8$ with $\Phi$ and $[\Phi]$, the superpotential is simplified to
\begin{equation}
W(\Phi, \phi_1, \phi_2) = \mu_0 [\Phi \Phi]
                          + \mu_3 \phi_1 \phi_2
                          + \begin{pmatrix}
                            \phi_1 & \phi_2
                            \end{pmatrix}
                            \begin{pmatrix}
                            b_{11} & b_{12}\\
                            b_{21} & b_{22}
                            \end{pmatrix}
                            \begin{pmatrix}
                            [\Phi \Phi] - \frac{1}{3} [\Phi]^2\\
                            [\Phi]^2
                            \end{pmatrix}.
\end{equation}
With proper choices of parameters $b_{11}$, $b_{12}$, $b_{21}$, $b_{22}$, $\mu_0$ and $\mu_3$, we expect that a non-zero Hermitian matrix-valued VEV of $\Phi$ can be obtained to give the SM fermion masses through Eq.~\eqref{eq:1-05}.

We will look for SUSY vacua which satisfy the F-flatness condition.  According to the last two subequations of Eq.~\eqref{eq:1-04}, the VEV's of the first derivatives of $W$ with respect to $\Phi$, $\phi_1$ and $\phi_2$ must vanish:
\begin{align}
0 & = \partial_\Phi W
    = 2 \mu_0 \Phi
      + 2 (b_{11} \phi_1 + b_{21} \phi_2) \Phi
      + 2 \left ( \left ( b_{12} - \frac{1}{3} b_{11} \right ) \phi_1
                  + \left ( b_{22} - \frac{1}{3} b_{21} \right ) \phi_2
          \right ) [\Phi] \mathbf{I}_{3 \times 3}, \label{eq:3-01}\\
0 & = \partial_{\phi_1} W
    = \mu_3 \phi_2 
      + b_{11} [\Phi \Phi]
      + \left ( b_{12} - \frac{1}{3} b_{11} \right ) [\Phi]^2, \label{eq:3-02}\\
0 & = \partial_{\phi_2} W
    = \mu_3 \phi_1
      + b_{21} [\Phi \Phi]
      + \left ( b_{22} - \frac{1}{3} b_{21} \right ) [\Phi]^2. \label{eq:3-03}
\end{align}
Eq.~\eqref{eq:3-02} and Eq.~\eqref{eq:3-03} give
\begin{align}
\phi_1 &= - \frac{1}{\mu_3}
            \left ( b_{21} [\Phi \Phi]
                    + \left ( b_{22} - \frac{1}{3} b_{21} \right ) [\Phi]^2
            \right ),\\
\phi_2 &= - \frac{1}{\mu_3}
            \left ( b_{11} [\Phi \Phi]
                    + \left ( b_{12} - \frac{1}{3} b_{11} \right ) [\Phi]^2
            \right ).
\end{align}
Replacing $\phi_1$ and $\phi_2$ with these expressions involving $[\Phi]$ and $[\Phi \Phi]$, Eq.~\eqref{eq:3-01} becomes 
\begin{equation}
\begin{split}
0 = \mbox{} & 2 
              \left ( \mu_0
                      - \frac{1}{\mu_3}
                        \left ( 2 b_{11} b_{21} [\Phi \Phi]
                                + \left ( b_{11} b_{22}
                                          + b_{12} b_{21}
                                          - \frac{2}{3} b_{11} b_{21}
                                  \right ) [\Phi]^2
                        \right )
              \right ) \Phi\\
            & - \frac{2}{\mu_3}
                \left ( \left ( b_{11} b_{22}
                                + b_{12} b_{21}
                                - \frac{2}{3} b_{11} b_{21}
                        \right ) [\Phi \Phi] \right.\\
            & \hphantom{- \frac{2}{\mu_3} \left ( \vphantom{\frac{2}{3}} \right.}
                \left. \mbox{}
                        + \left ( \frac{2}{9} b_{11} b_{21}
                                  + 2 b_{12} b_{22}
                                  - \frac{2}{3} b_{11} b_{22}
                                  - \frac{2}{3} b_{12} b_{21}
                          \right ) [\Phi]^2
                \right ) [\Phi] \mathbf{I}_{3 \times 3}. \label{eq:3-04}
\end{split}
\end{equation}

The VEV's of $\Phi$ and $\mathbf{I}_{3 \times 3}$ are linearly independent for a non-zero VEV of $\Phi$ without fine-tuning.  So their corresponding coefficients in Eq.~\eqref{eq:3-04} must vanish:
\begin{align}
0 &= \mu_0
     - \frac{1}{\mu_3}
       \left ( 2 b_{11} b_{21} [\Phi \Phi]
               + \left ( b_{11} b_{22}
                         + b_{12} b_{21}
                         - \frac{2}{3} b_{11} b_{21}
                 \right ) [\Phi]^2
       \right ), \label{eq:3-05}\\
0 &= \left ( b_{11} b_{22}
             + b_{12} b_{21}
             - \frac{2}{3} b_{11} b_{21}
     \right ) [\Phi \Phi]
     + \left ( \frac{2}{9} b_{11} b_{21}
               + 2 b_{12} b_{22}
               - \frac{2}{3} b_{11} b_{22}
               - \frac{2}{3} b_{12} b_{21}
       \right ) [\Phi]^2. \label{eq:3-06}
\end{align}
With the redefinition of parameters
\begin{equation}
a_{02} = b_{11} b_{21}, \quad
a_2 = b_{11} b_{22} + b_{12} b_{21}, \quad
a_4 = b_{12} b_{22}, \label{eq:3-07}
\end{equation}
Eq.~\eqref{eq:3-06} gives the modified version of the Koide formula:
\begin{equation}
K = \frac{[\Phi \Phi]}{[\Phi]^2}
  = \frac{2}{3} \times \frac{a_{02} / 3 - a_2 + 3 a_4}{2 a_{02} / 3 - a_2}
  = \frac{2}{3} \times
    \left ( 1 - \frac{a_{02} - 9 a_4}{2 a_{02} - 3 a_2} \right ). \label{eq:3-08}
\end{equation}
And Eq.~\eqref{eq:3-05} gives
\begin{equation}
\mu_0 \mu_3 = 2 a_{02} [\Phi \Phi]
              - \left ( \frac{2}{3} a_{02} - a_2 \right ) [\Phi]^2
            = \left ( 2 a_{02} \left ( K - \frac{1}{3} \right ) + a_2 \right ) [\Phi]^2. \label{eq:3-09}
\end{equation}
The coefficient of $[\Phi]^2$ is zero only if parameters are fine-tuned to satisfy $a_2^2 = 4 a_{02} a_4$.  Without such fine-tuning, a non-zero VEV of $\Phi$ requires non-zero $\mu_0$ and $\mu_3$.

The modified formula Eq.~\eqref{eq:3-08} contains three free parameters $a_{02}$, $a_2$ and $a_4$ after the parameter redefinition Eq.~\eqref{eq:3-07}.  Since $K$ only depends on ratios of the parameters, there are only two effective parameters in the modified formula.  Note that the previous formula Eq.~\eqref{eq:2-08} from the scalar potential model is identical to the formula Eq.~\eqref{eq:3-09} from the Yukawaon model if we make the identification from $a_0 / 2 + a_{02}$ to $a_{02}$.  But the formula is more naturally realized from the Yukawaon model without the tuning of $a_1$.

\section{Concluding remarks} \label{sec:4}

In this work, we derived Eq.~\eqref{eq:2-08} or Eq.~\eqref{eq:3-08}, the modified version of the Koide formula, from a flavor nonet scalar field $\Phi$ in either a scalar potential model or a Yukawaon model, with all terms respecting symmetries included in the scalar potential $V(\Phi)$ or the superpotential $W(\Phi)$.  In the scalar potential model, a $\mathbb{Z}_2$ symmetry is imposed in addition to the $\mathrm{SU}(3)$ flavor symmetry.  Linear and cubic terms in $V$ are then eliminated by the $\mathbb{Z}_2$ symmetry.  Quadratic terms are assumed to have a Higgs-like form with a negative mass-square $- \mu^2$, and one coefficient $a_1$ must vanish in order to have a non-zero VEV of $\Phi$.  From Eq.~\eqref{eq:2-09} we see that without fine-tuning, the magnitude of the VEV of $\Phi$ is proportional to $\mu$.  In the Yukawaon model, an R-symmetry is imposed in addition to the $\mathrm{SU}(3)$ flavor symmetry.  The nonet scalar field $\Phi$ is promoted to a nonet chiral superfield, but later assumed to get a non-zero Hermitian matrix-valued VEV\@.  The coefficients $a_0$ and $a_1$ in the scalar potential model naturally disappear in the Yukawaon model.  From Eq.~\eqref{eq:3-09} we see that without fine-tuning, the magnitude of the VEV of $\Phi$ is proportional to $\sqrt{\mu_0 \mu_3}$.  In both models, the modified formula is obtained with only two effective parameters.  The fact that $\mu_0$ characterizes R-symmetry breaking in the Yukawaon model indicates that there may be some deep relation between the SM Yukawa coupling terms and R-symmetry breaking dynamics in the hidden sector, which is worth to explore in the future.

As mentioned in~\cite{Koide:2018gdm}, $V_2$ or $W_2$ is introduced as a special combination of two irreducible terms proportional to $[\Phi_8 \Phi_8]$ and $[\Phi]^2$ respectively.  In the scalar potential model, the two terms have different renormalization group flow, which can modify the VEV of $\Phi$ and invalidates our derivation of the modified formula.  In the Yukawaon model, the superpotential does not get radiative corrections because of the non-renormalization theorem in SUSY.  Thus the VEV of $\Phi$ and the modified formula are protected from radiative corrections.

The two effective parameters $a_{02} / a_2$ and $a_4 / a_2$ in the modified formula Eq.~\eqref{eq:3-09} modify the Koide's character from its original value $K = 2 / 3$ when these parameters take non-zero values.  Such modification may serve as an alternative way, compared to previous approaches using gauged flavor symmetries~\cite{Sumino:2008hu, Sumino:2008hy, Sumino:2009bt, Koide:2012kn, Koide:2014doa, Koide:2016bte, Koide:2016qeq}, to cancel the possible QED correction of $K$.  The modified value of $K$ can also fit all possible values of $K$ for charged leptons, up quarks and down quarks.  For an arbitrary set of fermion masses, the Cauchy-Schwarz inequality and the positiveness of masses leads to
\begin{equation}
\frac{1}{3} \le K
            \le 1,
\end{equation}
which gives the parameter range
\begin{equation}
- \frac{1}{2} \le \frac{a_{02} - 9 a_4}{2 a_{02} - 3 a_2}
              \le \frac{1}{2}.
\end{equation}
The value $K = 2 / 3$ for charged leptons corresponds to $a_{02} = 9 a_4$, which covers $a_{02} = a_4 = 0$ as a special case, corresponding to the superpotential
\begin{equation}
W = \mu_0 [\Phi \Phi] + \mu_3 \phi_1 \phi_2 + b_{11} \phi_1 [\Phi_8 \Phi_8] + b_{22} \phi_2 [\Phi]^2
\end{equation}
used previously in~\cite{Koide:2018gdm}.

When applying the formula to quarks, it should be noted that quark masses can not be directly measured, because quarks are confined inside hadrons.  The values of quark masses are renormalization scheme and scale dependent.  Fortunately, the Koide's character $K$ depends only on mass ratios of fermions.  Since QCD renormalization factors are identical for all quarks if the same scheme and scale are used, in mass ratios all these factors cancel exactly.  Quark mass ratios are also calculated in lattice QCD with higher precision than individual quark masses~\cite{FlavourLatticeAveragingGroup:2019iem}.  Thus we quote the data of quark mass ratios~\cite{Zyla:2020zbs}
\begin{align}
m_u / m_d &= 0.47^{+ 0.06}_{- 0.07},\\
m_s / ((m_u + m_d) / 2) &= 27.3^{+ 0.7}_{- 1.3},\\
m_c / m_s &= 11.72 \pm 0.25,\\
m_b / m_s &= 53.94 \pm 0.12
\end{align}
with their 1-$\sigma$ errors.  There is no lattice QCD calculation for mass ratios involving $m_t$.  Using the data of the on-shell mass for the top quark and the modified minimal subtraction ($\overline{\text{MS}}$) mass for the bottom quark~\cite{Zyla:2020zbs}:
\begin{align}
M_t &= 172.76 \pm 0.30 \ \text{GeV} / c^2,\\
m_b(m_b) &= 4.18^{+ 0.03}_{- 0.02} \ \text{GeV} / c^2,
\end{align}
the scale dependence of running masses can be calculated from renormalization group equations using the \texttt{RunDec} or \texttt{CRunDec} package~\cite{Chetyrkin:2000yt, Schmidt:2012az, Herren:2017osy}.  The resulting top quark $\overline{\text{MS}}$ mass $m_t(m_t)$ and the bottom quark running mass at the scale $\mu = m_t(m_t)$ are
\begin{align}
m_t(m_t) &= 163.39 \pm 0.29 \ \text{GeV} / c^2,\\
m_b(m_t) &= 2.737^{+ 0.015}_{- 0.022} \ \text{GeV} / c^2,
\end{align}
which give the mass ratio
\begin{equation}
m_t / m_b = 59.70^{+ 0.58}_{- 0.42}
\end{equation}
with its 1-$\sigma$ error.  These mass ratios lead to the Koide's character
\begin{align}
  K_{\text{up quarks}} &= \frac{m_u + m_c + m_t}{\left ( \sqrt{m_u} + \sqrt{m_c} + \sqrt{m_t} \right )^2}
                        = 0.8882^{+ 0.0019}_{- 0.0018}
                        = \frac{2}{3} \times (1.3322^{+ 0.0029}_{- 0.0027}),\\
K_{\text{down quarks}} &= \frac{m_d + m_s + m_b}{\left ( \sqrt{m_d} + \sqrt{m_s} + \sqrt{m_b} \right )^2}
                        = 0.7491^{+ 0.0015}_{- 0.0021}
                        = \frac{2}{3} \times (1.1237^{+ 0.0022}_{- 0.0032}).
\end{align}
The mean values of $K$ can be fitted with $a_2 = 1.670 a_{02} - 9.029 a_4$ for up quarks and $a_2 = 3.361 a_{02} - 24.25 a_4$ for down quarks.  We expect that neutrino masses may also be interpreted by considering a see-saw mechanism together with the Yukawaon model, and the experimental verification depends on whether neutrino masses are in the normal or inverted hierarchy.  Thus our scalar potential or superpotential constructed from symmetries may provide a natural interpretation of the SM fermion mass spectrum.  It is still a challenge to explore possible symmetry settings which may give the correct values of $a_{02}$, $a_2$ and $a_4$ for the mass formula in each sector.

\section*{Acknowledgement}

The authors thank Yoshio Koide, Jinmian Li, Bo-Qiang Ma and Zhong-Qi Ma for helpful discussions.  This work is supported by the National Natural Science Foundation of China under grant 11305110.


\begin{thebibliography}{50}

\bibitem{Koide:1982si}
Y.~Koide,
``Fermion - Boson Two-body Model of Quarks and Leptons and Cabibbo Mixing,''
Lett.\ Nuovo Cim.\ \textbf{34} (1982), 201
doi:10.1007/BF02817096

\bibitem{Koide:1983a}
Y.~Koide,
``A fermion-boson composite model of quarks and leptons,''
Phys.\ Lett.\ B \textbf{120} (1983), 161
doi:10.1016/0370-2693(83)90644-5

\bibitem{Koide:1983b}
Y.~Koide,
``New view of quark and lepton mass hierarchy,''
Phys.\ Rev.\ D \textbf{28} (1983), 252
doi:10.1103/PhysRevD.28.252

\bibitem{Patrignani:2016xqp}
C.~Patrignani \textit{et al.} [Particle Data Group],
``Review of Particle Physics,''
Chin.\ Phys.\ C \textbf{40} (2016) no.10, 100001
doi:10.1088/1674-1137/40/10/100001

\bibitem{Tanabashi:2018oca}
M.~Tanabashi \textit{et al.} [Particle Data Group],
``Review of Particle Physics,''
Phys.\ Rev.\ D \textbf{98} (2018) no.3, 030001
doi:10.1103/PhysRevD.98.030001

\bibitem{Zyla:2020zbs}
P.~A.~Zyla \textit{et al.} [Particle Data Group],
``Review of Particle Physics,''
PTEP \textbf{2020} (2020) no.8, 083C01
doi:10.1093/ptep/ptaa104

\bibitem{Li:2006et}
N.~Li and B.~Q.~Ma,
``Energy scale independence of Koide's relation for quark and lepton masses,''
Phys.\ Rev.\ D \textbf{73} (2006), 013009
doi:10.1103/PhysRevD.73.013009
[arXiv:hep-ph/0601031 [hep-ph]].

\bibitem{Xing:2006vk}
Z.~z.~Xing and H.~Zhang,
``On the Koide-like relations for the running masses of charged leptons, neutrinos and quarks,''
Phys.\ Lett.\ B \textbf{635} (2006), 107-111
doi:10.1016/j.physletb.2006.02.051
[arXiv:hep-ph/0602134 [hep-ph]].

\bibitem{Xing:2019vks}
Z.~z.~Xing,
``Flavor structures of charged fermions and massive neutrinos,''
Phys.\ Rept.\ \textbf{854} (2020), 1-147
doi:10.1016/j.physrep.2020.02.001
[arXiv:1909.09610 [hep-ph]].

\bibitem{Sumino:2008hu}
Y.~Sumino,
``Family Gauge Symmetry and Koide's Mass Formula,''
Phys.\ Lett.\ B \textbf{671} (2009), 477-480
doi:10.1016/j.physletb.2008.12.060
[arXiv:0812.2090 [hep-ph]].

\bibitem{Sumino:2008hy}
Y.~Sumino,
``Family Gauge Symmetry as an Origin of Koide's Mass Formula and Charged Lepton Spectrum,''
JHEP \textbf{05} (2009), 075
doi:10.1088/1126-6708/2009/05/075
[arXiv:0812.2103 [hep-ph]].

\bibitem{Sumino:2009bt}
Y.~Sumino,
``Family Gauge Symmetry as an Origin of Koide's Mass Formula and Charged Lepton Spectrum,''
[arXiv:0903.3640 [hep-ph]].

\bibitem{Koide:2012kn}
Y.~Koide and T.~Yamashita,
``Family Gauge Bosons with an Inverted Mass Hierarchy,''
Phys.\ Lett.\ B \textbf{711} (2012), 384-389
doi:10.1016/j.physletb.2012.04.028
[arXiv:1203.2028 [hep-ph]].

\bibitem{Koide:2014doa}
Y.~Koide,
``Spectroscopy of Family Gauge Bosons,''
Phys.\ Lett.\ B \textbf{736} (2014), 499-505
doi:10.1016/j.physletb.2014.07.061
[arXiv:1405.6778 [hep-ph]].

\bibitem{Koide:2016bte}
Y.~Koide and M.~Yamanaka,
``Muon\textendash{}electron conversion in a family gauge boson model,''
Phys.\ Lett.\ B \textbf{762} (2016), 41-46
doi:10.1016/j.physletb.2016.09.004
[arXiv:1608.01650 [hep-ph]].

\bibitem{Koide:2016qeq}
Y.~Koide,
``Sumino\textquoteright{}s cancellation mechanism in an anomaly-free model,''
Mod.\ Phys.\ Lett.\ A \textbf{32} (2017) no.1, 1750062
doi:10.1142/S0217732317500626
[arXiv:1608.04514 [hep-ph]].

\bibitem{Foot:1994yn}
R.~Foot,
``A Note on Koide's lepton mass relation,''
[arXiv:hep-ph/9402242 [hep-ph]].

\bibitem{Esposito:1995bw}
S.~Esposito and P.~Santorelli,
``A Geometric picture for fermion masses,''
Mod.\ Phys.\ Lett.\ A \textbf{10} (1995), 3077-3082
doi:10.1142/S0217732395003215
[arXiv:hep-ph/9603369 [hep-ph]].

\bibitem{Li:2005rp}
N.~Li and B.~Q.~Ma,
``Estimate of neutrino masses from Koide's relation,''
Phys.\ Lett.\ B \textbf{609} (2005), 309-316
doi:10.1016/j.physletb.2005.01.066
[arXiv:hep-ph/0505028 [hep-ph]].

\bibitem{Rivero:2005vj}
A.~Rivero and A.~Gsponer,
``The Strange formula of Dr. Koide,''
[arXiv:hep-ph/0505220 [hep-ph]].

\bibitem{Gerard:2005ad}
J.~M.~Gerard, F.~Goffinet and M.~Herquet,
``A New look at an old mass relation,''
Phys.\ Lett.\ B \textbf{633} (2006), 563-566
doi:10.1016/j.physletb.2005.12.054
[arXiv:hep-ph/0510289 [hep-ph]].

\bibitem{Guo:2007rn}
Z.~Q.~Guo and B.~Q.~Ma,
``Determining quark and lepton mass matrices by a geometrical interpretation,''
Phys.\ Lett.\ B \textbf{647} (2007), 436-445
doi:10.1016/j.physletb.2007.02.031
[arXiv:hep-ph/0702288 [hep-ph]].

\bibitem{Rodejohann:2011jj}
W.~Rodejohann and H.~Zhang,
``Extension of an empirical charged lepton mass relation to the neutrino sector,''
Phys.\ Lett.\ B \textbf{698} (2011), 152-156
doi:10.1016/j.physletb.2011.03.007
[arXiv:1101.5525 [hep-ph]].

\bibitem{Kartavtsev:2011jt}
A.~Kartavtsev,
``A Remark on the Koide relation for quarks,''
[arXiv:1111.0480 [hep-ph]].

\bibitem{Cao:2012un}
F.~G.~Cao,
``Neutrino masses from lepton and quark mass relations and neutrino oscillations,''
Phys.\ Rev.\ D \textbf{85} (2012), 113003
doi:10.1103/PhysRevD.85.113003
[arXiv:1205.4068 [hep-ph]].

\bibitem{Zenczykowski:2012fg}
P.~Zenczykowski,
``Remark on Koide's Z3-symmetric parametrization of quark masses,''
Phys.\ Rev.\ D \textbf{86} (2012), 117303
doi:10.1103/PhysRevD.86.117303
[arXiv:1210.4125 [hep-ph]].

\bibitem{Zenczykowski:2013bb}
P.~\.Zenczykowski,
``Koide\textquoteright{}s $Z_3$-symmetric parametrization, quark masses, and mixings,''
Phys.\ Rev.\ D \textbf{87} (2013) no.7, 077302
doi:10.1103/PhysRevD.87.077302
[arXiv:1301.4143 [hep-ph]].

\bibitem{Gao:2015xnv}
G.~H.~Gao and N.~Li,
``Explorations of two empirical formulas for fermion masses,''
Eur.\ Phys.\ J.\ C \textbf{76} (2016) no.3, 140
doi:10.1140/epjc/s10052-016-3990-3
[arXiv:1512.06349 [hep-ph]].

\bibitem{Huang:2016ocs}
Y.~C.~Huang, S.~T.~Iqbal, Z.~Lei and W.~Y.~Wang,
``Some new symmetric relations and prediction of left- and right-handed neutrino masses using Koide\textquoteright{}s relation,''
Chin.\ Phys.\ C \textbf{41} (2017) no.4, 043104
doi:10.1088/1674-1137/41/4/043104
[arXiv:1601.00754 [hep-ph]].

\bibitem{Huang:2016sht}
Y.~C.~Huang, S.~T.~Iqbal and W.~Y.~Wang,
``New symmetric relations among lepton masses and prediction of lepton masses of the fourth generation,''
Mod.\ Phys.\ Lett.\ A \textbf{31} (2016) no.18, 1650115
doi:10.1142/S0217732316501157

\bibitem{Froggatt:1978nt}
C.~D.~Froggatt and H.~B.~Nielsen,
``Hierarchy of Quark Masses, Cabibbo Angles and CP Violation,''
Nucl.\ Phys.\ B \textbf{147} (1979), 277-298
doi:10.1016/0550-3213(79)90316-X

\bibitem{Koide:1989ds}
Y.~Koide,
``Charged Lepton Mass Matrix With Democratic Family Mixing,''
Z.\ Phys.\ C \textbf{45} (1989), 39
doi:10.1007/BF01556669

\bibitem{Koide:1989jq}
Y.~Koide,
``Charged lepton mass sum rule from U(3) family Higgs potential model,''
Mod.\ Phys.\ Lett.\ A \textbf{5} (1990), 2319-2324
doi:10.1142/S0217732390002663

\bibitem{Koide:1992vs}
Y.~Koide,
``Should the renewed tau mass value 1777-MeV be taken seriously?,''
Mod.\ Phys.\ Lett.\ A \textbf{8} (1993), 2071-2078
doi:10.1142/S0217732393001781

\bibitem{Koide:1993da}
Y.~Koide,
``Phenomenological quark mass matrix model with two adjustable parameters,''
Phys.\ Rev.\ D \textbf{49} (1994), 2638-2641
doi:10.1103/PhysRevD.49.2638
[arXiv:hep-ph/9309330 [hep-ph]].

\bibitem{Koide:1994wu}
Y.~Koide and H.~Fusaoka,
``Seesaw type quark and lepton mass matrices and SU(3) family nonet Higgs bosons,''
[arXiv:hep-ph/9403354 [hep-ph]].

\bibitem{Koide:1995ie}
Y.~Koide,
``New physics from U(3) family nonet Higgs boson scenario,''
[arXiv:hep-ph/9501408 [hep-ph]].

\bibitem{Koide:1995xk}
Y.~Koide and M.~Tanimoto,
``U(3) family nonet Higgs boson and its phenomenology,''
Z.\ Phys.\ C \textbf{72} (1996), 333-344
doi:10.1007/BF02909162
[arXiv:hep-ph/9505333 [hep-ph]].

\bibitem{Koide:1999mx}
Y.~Koide,
``Universal seesaw mass matrix model with an S(3) symmetry,''
Phys.\ Rev.\ D \textbf{60} (1999), 077301
doi:10.1103/PhysRevD.60.077301
[arXiv:hep-ph/9905416 [hep-ph]].

\bibitem{Koide:2000zi}
Y.~Koide,
``Quark and lepton mass matrices with a cyclic permutation invariant form,''
[arXiv:hep-ph/0005137 [hep-ph]].

\bibitem{Koide:2005nv}
Y.~Koide,
``Challenge to the mystery of the charged lepton mass formula,''
[arXiv:hep-ph/0506247 [hep-ph]].

\bibitem{Koide:2005za}
Y.~Koide,
``Seesaw mass matrix model of quarks and leptons with flavor-triplet Higgs scalars,''
Eur.\ Phys.\ J.\ C \textbf{48} (2006), 223-228
doi:10.1140/epjc/s10052-006-0009-5
[arXiv:hep-ph/0508301 [hep-ph]].

\bibitem{Koide:2005ep}
Y.~Koide,
``Permutation symmetry S(3) and VEV structure of flavor-triplet Higgs scalars,''
Phys.\ Rev.\ D \textbf{73} (2006), 057901
doi:10.1103/PhysRevD.73.057901
[arXiv:hep-ph/0509214 [hep-ph]].

\bibitem{Mohapatra:2006gs}
R.~N.~Mohapatra and A.~Y.~Smirnov,
``Neutrino Mass and New Physics,''
Ann.\ Rev.\ Nucl.\ Part.\ Sci.\ \textbf{56} (2006), 569-628
doi:10.1146/annurev.nucl.56.080805.140534
[arXiv:hep-ph/0603118 [hep-ph]].

\bibitem{Koide:2006dn}
Y.~Koide,
``Tribimaximal Neutrino Mixing and a Relation Between Neutrino- and Charged Lepton-Mass Spectra,''
J.\ Phys.\ G \textbf{34} (2007), 1653-1664
doi:10.1088/0954-3899/34/7/006
[arXiv:hep-ph/0605074 [hep-ph]].

\bibitem{Ma:2006ht}
E.~Ma,
``Lepton Family Symmetry and Possible Application to the Koide Mass Formula,''
Phys.\ Lett.\ B \textbf{649} (2007), 287-291
doi:10.1016/j.physletb.2007.04.020
[arXiv:hep-ph/0612022 [hep-ph]].

\bibitem{Koide:2006vs}
Y.~Koide,
``S(3) symmetry and neutrino masses and mixings,''
Eur.\ Phys.\ J.\ C \textbf{50} (2007), 809-816
doi:10.1140/epjc/s10052-007-0261-3
[arXiv:hep-ph/0612058 [hep-ph]].

\bibitem{Koide:2007kw}
Y.~Koide,
``A(4) symmetry and lepton masses and mixing,''
Eur.\ Phys.\ J.\ C \textbf{52} (2007), 617-623
doi:10.1140/epjc/s10052-007-0433-1
[arXiv:hep-ph/0701018 [hep-ph]].

\bibitem{Koide:2007sr}
Y.~Koide,
``S(4) flavor symmetry embedded into SU(3) and lepton masses and mixing,''
JHEP \textbf{08} (2007), 086
doi:10.1088/1126-6708/2007/08/086
[arXiv:0705.2275 [hep-ph]].

\bibitem{Koide:2007eu}
Y.~Koide,
``Charged Lepton Mass Formula: Development and Prospect,''
Int.\ J.\ Mod.\ Phys.\ E \textbf{16} (2007), 1417-1426
doi:10.1142/S0218301307006770
[arXiv:0706.2534 [hep-ph]].

\bibitem{Koide:2010vu}
Y.~Koide,
``Charged Lepton Mass Spectrum and a Scalar Potential Model,''
Phys.\ Rev.\ D \textbf{81} (2010), 097901
doi:10.1103/PhysRevD.81.097901
[arXiv:1004.0580 [hep-ph]].

\bibitem{Haba:2008wr}
N.~Haba and Y.~Koide,
``F-term Induced Flavor Mass Spectrum,''
JHEP \textbf{06} (2008), 023
doi:10.1088/1126-6708/2008/06/023
[arXiv:0801.3301 [hep-ph]].

\bibitem{Koide:2008eq}
Y.~Koide,
``How to Evade a No-Go Theorem in Flavor Symmetries,''
AIP Conf.\ Proc.\ \textbf{1015} (2008) no.1, 80-86
doi:10.1063/1.2939063
[arXiv:0801.3491 [hep-ph]].

\bibitem{Koide:2008ey}
Y.~Koide,
``U(3)-Flavor Nonet Scalar as an Origin of the Flavor Mass Spectra,''
Phys.\ Lett.\ B \textbf{662} (2008), 43-48
doi:10.1016/j.physletb.2008.02.059
[arXiv:0802.1084 [hep-ph]].

\bibitem{Koide:2008tr}
Y.~Koide,
``Charged Lepton Mass Relations in a Supersymmetric Yukawaon Model,''
Phys.\ Rev.\ D \textbf{79} (2009), 033009
doi:10.1103/PhysRevD.79.033009
[arXiv:0811.3470 [hep-ph]].

\bibitem{Koide:2010zz}
Y.~Koide,
``Can massless and light Yukawaons be harmless?,''
Int.\ J.\ Mod.\ Phys.\ A \textbf{25} (2010), 1725-1738
doi:10.1142/S0217751X10048068
[arXiv:0902.4501 [hep-ph]].

\bibitem{Koide:2008qm}
Y.~Koide,
``Phenomenological Meaning of a Neutrino Mass Matrix Related to Up-Quark Masses,''
Phys.\ Rev.\ D \textbf{78} (2008), 093006
doi:10.1103/PhysRevD.78.093006
[arXiv:0809.2449 [hep-ph]].

\bibitem{Koide:2013eca}
Y.~Koide and H.~Nishiura,
``Yukawaon Model with Anomaly Free Set of Quarks and Leptons in a U(3) Family Symmetry,''
Phys.\ Rev.\ D \textbf{88} (2013) no.11, 116004
doi:10.1103/PhysRevD.88.116004
[arXiv:1308.2129 [hep-ph]].

\bibitem{Koide:2008sj}
Y.~Koide,
``Empirical Neutrino Mass Matrix Related to Up-Quark Masses,''
J.\ Phys.\ G \textbf{35} (2008), 125004
doi:10.1088/0954-3899/35/12/125004
[arXiv:0803.3101 [hep-ph]].

\bibitem{Koide:2008he}
Y.~Koide,
``O(3) flavor symmetry and an empirical neutrino mass matrix,''
Phys.\ Lett.\ B \textbf{665} (2008), 227-230
doi:10.1016/j.physletb.2008.06.027
[arXiv:0804.4267 [hep-ph]].

\bibitem{Koide:2008kw}
Y.~Koide,
``Neutrino Mass Matrix Related to Up-Quark Masses and Nearly Tribimaximal Mixing: Based on a Yukawaon model,''
Int.\ J.\ Mod.\ Phys.\ A \textbf{24} (2009), 3469-3475
doi:10.1142/S0217751X09047077
[arXiv:0812.3203 [hep-ph]].

\bibitem{Koide:2009zz}
Y.~Koide,
``Yukawaon model in the quark sector and nearly tribimaximal neutrino mixing,''
Phys.\ Lett.\ B \textbf{680} (2009), 76-80
doi:10.1016/j.physletb.2009.08.038
[arXiv:0904.1644 [hep-ph]].

\bibitem{Koide:2010np}
Y.~Koide,
``Yukawaon Approach to the Sumino Relation for Charged Lepton Masses,''
Phys.\ Lett.\ B \textbf{687} (2010), 219-224
doi:10.1016/j.physletb.2010.03.019
[arXiv:1001.4877 [hep-ph]].

\bibitem{Koide:2010hp}
Y.~Koide,
``Yukawaon Model with U(3)$\times$O(3) Family Symmetries,''
J.\ Phys.\ G \textbf{38} (2011), 085004
doi:10.1088/0954-3899/38/8/085004
[arXiv:1011.1064 [hep-ph]].

\bibitem{Nishiura:2010rt}
H.~Nishiura and Y.~Koide,
``Unified description of quark and lepton mixing matrices based on a yukawaon model,''
Phys.\ Rev.\ D \textbf{83} (2011), 035010
doi:10.1103/PhysRevD.83.035010
[arXiv:1011.1312 [hep-ph]].

\bibitem{Koide:2012zz}
Y.~Koide,
``SU(5)-Compatible Yukawaon Model,''
Int.\ J.\ Mod.\ Phys.\ A \textbf{27} (2012), 1250028
doi:10.1142/S0217751X12500285
[arXiv:1106.0971 [hep-ph]].

\bibitem{Koide:2011wj}
Y.~Koide and H.~Nishiura,
``Neutrino Mass Matrix with No Adjustable Parameters,''
Eur.\ Phys.\ J.\ C \textbf{72} (2012), 1933
doi:10.1140/epjc/s10052-012-1933-1
[arXiv:1106.5202 [hep-ph]].

\bibitem{Koide:2012fw}
Y.~Koide and H.~Nishiura,
``Yukawaon Model with U(3)$\times$S$_3$ Family Symmetries,''
Phys.\ Lett.\ B \textbf{712} (2012), 396-400
doi:10.1016/j.physletb.2012.05.014
[arXiv:1202.5815 [hep-ph]].

\bibitem{Koide:2012ji}
Y.~Koide and H.~Nishiura,
``Large $\theta_{13}^\nu$ and Unified Description of Quark and Lepton Mixing Matrices,''
Eur.\ Phys.\ J.\ C \textbf{73} (2013) no.1, 2277
doi:10.1140/epjc/s10052-013-2277-1
[arXiv:1209.1275 [hep-ph]].

\bibitem{Koide:2013ie}
Y.~Koide and H.~Nishiura,
``Neutrino Mass Matrix Model with a Bilinear Form,''
JHEP \textbf{04} (2013), 166
doi:10.1007/JHEP04(2013)166
[arXiv:1301.4312 [hep-ph]].

\bibitem{Koide:2014nxa}
Y.~Koide and H.~Nishiura,
``Universal Bilinear Form of Quark and Lepton Mass Matrices,''
Phys.\ Rev.\ D \textbf{90} (2014) no.1, 016009
doi:10.1103/PhysRevD.90.016009
[arXiv:1405.0069 [hep-ph]].

\bibitem{Koide:2014oxa}
Y.~Koide and H.~Nishiura,
``Leptonic $CP$ violating Phase in the Yukawaon Model,''
Phys.\ Rev.\ D \textbf{90} (2014) no.11, 117903
[erratum: Phys.\ Rev.\ D \textbf{91} (2015) no.11, 119904]
doi:10.1103/PhysRevD.90.117903
[arXiv:1410.8653 [hep-ph]].

\bibitem{Koide:2015ura}
Y.~Koide and H.~Nishiura,
``Origin of Hierarchical Structures of Quark and Lepton Mass Matrices,''
Phys.\ Rev.\ D \textbf{91} (2015) no.11, 116002
doi:10.1103/PhysRevD.91.116002
[arXiv:1503.04900 [hep-ph]].

\bibitem{Koide:2015hya}
Y.~Koide and H.~Nishiura,
``Quark and lepton mass matrix model with only six family-independent parameters,''
Phys.\ Rev.\ D \textbf{92} (2015) no.11, 111301
doi:10.1103/PhysRevD.92.111301
[arXiv:1510.05370 [hep-ph]].

\bibitem{Koide:2015ype}
Y.~Koide and H.~Nishiura,
``Quark and Lepton Mass Matrices Described by Charged Lepton Masses,''
Mod.\ Phys.\ Lett.\ A \textbf{31} (2016) no.20, 1650125
doi:10.1142/S021773231650125X
[arXiv:1512.08386 [hep-ph]].

\bibitem{Koide:2017lan}
Y.~Koide and H.~Nishiura,
``Flavon VEV Scales in U(3)$\times$U(3)$'$ Model,''
Int.\ J.\ Mod.\ Phys.\ A \textbf{32} (2017) no.15, 1750085
doi:10.1142/S0217751X17500853
[arXiv:1701.06287 [hep-ph]].

\bibitem{Koide:2018fsj}
Y.~Koide and H.~Nishiura,
``Parameter-Independent Quark Mass Relation in the U(3)$\times$U(3)$'$ Model,''
Mod.\ Phys.\ Lett.\ A \textbf{33} (2018) no.39, 1850230
doi:10.1142/S0217732318502309
[arXiv:1805.07334 [hep-ph]].

\bibitem{Koide:2017lrf}
Y.~Koide,
``Another Formula for the Charged Lepton Masses,''
Phys.\ Lett.\ B \textbf{777} (2018), 131-133
doi:10.1016/j.physletb.2017.12.004
[arXiv:1711.03221 [hep-ph]].

\bibitem{Koide:2009hn}
Y.~Koide,
``Charged Lepton Mass Spectrum and Supersymmetric Yukawaon Model,''
Phys.\ Lett.\ B \textbf{681} (2009), 68-73
doi:10.1016/j.physletb.2009.09.065
[arXiv:0906.3370 [hep-ph]].

\bibitem{Koide:2018gdm}
Y.~Koide and T.~Yamashita,
``Charged Lepton Mass Relations in a SUSY Scenario,''
Phys.\ Lett.\ B \textbf{787} (2018), 171-174
doi:10.1016/j.physletb.2018.10.058
[arXiv:1805.09533 [hep-ph]].

\bibitem{Autonne:1915a}
L.~C.~Autonne,
``Sur les matrices hypohermitiennes et sur les matrices unitaires,''
Ann.\ Univ.\ Lyon \textbf{38} (1915), 1-77.

\bibitem{Takagi:1924a}
T.~Takagi,
``On an Algebraic Problem Reluted to an Analytic Theorem of Caratheodory and Fejer and on an Allied Theorem of Landau,''
Jpn.\ J.\ Math.\ Trans.\ Abs.\ \textbf{1} (1924), 83-93
doi:10.4099/jjm1924.1.0\_83.

\bibitem{FlavourLatticeAveragingGroup:2019iem}
S.~Aoki \textit{et al.} [Flavour Lattice Averaging Group],
``FLAG Review 2019: Flavour Lattice Averaging Group (FLAG),''
Eur.\ Phys.\ J.\ C \textbf{80} (2020) no.2, 113
doi:10.1140/epjc/s10052-019-7354-7
[arXiv:1902.08191 [hep-lat]].

\bibitem{Chetyrkin:2000yt}
K.~G.~Chetyrkin, J.~H.~Kuhn and M.~Steinhauser,
``RunDec: A Mathematica package for running and decoupling of the strong coupling and quark masses,''
Comput.\ Phys.\ Commun.\ \textbf{133} (2000), 43-65
doi:10.1016/S0010-4655(00)00155-7
[arXiv:hep-ph/0004189 [hep-ph]].

\bibitem{Schmidt:2012az}
B.~Schmidt and M.~Steinhauser,
``CRunDec: a C++ package for running and decoupling of the strong coupling and quark masses,''
Comput.\ Phys.\ Commun.\ \textbf{183} (2012), 1845-1848
doi:10.1016/j.cpc.2012.03.023
[arXiv:1201.6149 [hep-ph]].

\bibitem{Herren:2017osy}
F.~Herren and M.~Steinhauser,
``Version 3 of RunDec and CRunDec,''
Comput.\ Phys.\ Commun.\ \textbf{224} (2018), 333-345
doi:10.1016/j.cpc.2017.11.014
[arXiv:1703.03751 [hep-ph]].

\end{thebibliography}
\end{document}